\documentclass[twocolumn,showpacs,nofootinbib]{revtex4-1}
\usepackage{epsfig,amssymb,amsmath,latexsym}
\usepackage{pifont}
\usepackage{ulem}
\usepackage{hyperref}
\usepackage{subfigure}

\usepackage{graphicx}
\usepackage{dcolumn}
\usepackage{bm}
\usepackage{epsfig}
\usepackage{color}

\usepackage{color}
\usepackage{hyperref}

\hypersetup{
    colorlinks=true,
    linkcolor=red,
    citecolor=blue,
}

\definecolor{nicegreen}{rgb}{0.1,0.5,0.1}

\def\fun#1#2{\lower3.6pt\vbox{\baselineskip0pt\lineskip.9pt
  \ialign{$\mathsurround=0pt#1\hfil##\hfil$\crcr#2\crcr\sim\crcr}}}
\def\simgt{\mathrel{\lower0.6ex\hbox{$\buildrel {\textstyle >}
 \over {\scriptstyle \sim}$}}}
\def\simlt{\mathrel{\lower0.6ex\hbox{$\buildrel {\textstyle <}
 \over {\scriptstyle \sim}$}}}

\def\bea{\begin{eqnarray}}
\def\eea{\end{eqnarray}}
\def\be{\begin{equation}}
\def\ee{\end{equation}}

\input epsf

\def\be{\begin{equation}}
\def\ee{\end{equation}}
\def\ba{\begin{eqnarray}}
\def\ea{\end{eqnarray}}

\newcommand{\no}{\noindent}

\begin{document}

\preprint{}

\title{Neural Posterior Estimation with guaranteed exact coverage:\\ the ringdown of GW150914}

\author{Marco Crisostomi$^{1,2}$, Kallol Dey$^{3}$, Enrico Barausse$^{1,2}$, Roberto Trotta$^{1,2,4,5}$}
	
\affiliation{
$^1$SISSA, Via Bonomea 265, 34136 Trieste, Italy and INFN Sezione di Trieste \\
$^2$IFPU - Institute for Fundamental Physics of the Universe, Via Beirut 2, 34014 Trieste, Italy\\
$^3$School of Physics, Indian Institute of Science Education and Research Thiruvananthapuram,\\ Maruthamala PO, Vithura, Thiruvananthapuram 695551, Kerala, India \\
$^4$ Physics Department, Imperial College London, Prince Consort Rd, SW7 2AZ London, UK \\
$^5$Centro Nazionale ``High Performance Computer, 
Big Data and Quantum Computing''}

\date{\small \today}

\begin{abstract}
We analyze the ringdown phase of the first detected black-hole merger, GW150914, using a simulation-based inference pipeline based on masked autoregressive flows. We obtain approximate marginal posterior distributions for the ringdown parameters, namely the mass, spin, and the amplitude and phases of the dominant mode and its first overtone. Thanks to the locally amortized nature of our method, we are able to calibrate our posteriors with injected simulations, producing  posterior regions with guaranteed (i.e.~exact) frequentist coverage of the true values. For GW150914, our calibrated posteriors provide only mild evidence ($\sim 2\sigma$) for the presence of an overtone, even if the ringdown is assumed to start at the peak of the amplitude.
\end{abstract}


\maketitle

\section{Introduction}

\noindent

The first direct detection of gravitational waves (GWs) from the black hole (BH) merger GW150914~\cite{LIGOScientific:2016aoc} marked a pivotal moment in physics history. The later detections by the LIGO-Virgo-KAGRA (LVK) collaboration~\cite{LIGOScientific:2021djp} notwithstanding, GW150914 remains one of the most interesting events  thus far, mainly because of its relatively loud ringdown signal. The latter can be described by a linear superposition of damped sinusoids (or quasi-normal modes, QNMs), which characterize the GW emission from the perturbed BH remnant forming from the merger~\cite{Berti:2009kk}.
QNM frequencies and decay times depend only on the mass and spin of the remnant BH because of the no-hair theorem of general relativity~\cite{Kerr:1963ud, Newman:1965tw, Carter:1968rr, Robinson:1975bv}. Therefore, observation of one single QNM allows one to infer the mass and spin, and any additional observed QNM allows for performing consistency tests of the no-hair theorem~\cite{1980ApJ...239..292D,Dreyer:2003bv,Berti:2005ys}. These tests should be feasible with future detectors~\cite{Berti:2016lat}, but the presence of any QNMs
beyond the dominant fundamental mode 
in current data is hotly debated.

Indeed, while the presence of the fundamental ($n=0$, $\ell=m=2$) QNM in GW150914 data has been robustly established~\cite{LIGOScientific:2016lio}, inferences on the first overtone of that mode, i.e. the $n=1$, $\ell=m=2$ QNM, yield discrepant results.
Ref.~\cite{Isi:2019aib} claimed evidence at $3.6\sigma$ for this mode from GW150914 data (as expected from their previous analysis of numerical-relativity simulations~\cite{Giesler:2019uxc}), but this was not confirmed by the later Ref.~\cite{Cotesta:2022pci}. Indeed, the evidence for higher-damping modes in current data (be them the overtones of the $\ell=m=2$ mode, or the fundamental $\ell=m=3$ mode
whose presence in GW190521 is currently debated~\cite{Capano:2021etf, Capano:2022zqm, LIGOScientific:2021sio})
is highly sensitive to assumptions about detector noise, to the statistical methods used, to the choice of the ringdown starting time~\cite{Cotesta:2022pci}, and to the possible presence of nonlinear QNMs~\cite{Cheung:2022rbm, Mitman:2022qdl, Nee:2023osy}. These uncertainties have generated a lively debate~\cite{Isi:2022mhy,Baibhav:2023clw} between the authors of Refs.~\cite{Isi:2019aib,Cotesta:2022pci}, with more groups re-analyzing the GW150914 data with mixed results --
e.g. Refs.~\cite{Finch:2022ynt,CalderonBustillo:2020rmh}
find
only mild evidence ($1-2\sigma$) for the $n=1$ overtone (see also Ref.~\cite{LIGOScientific:2020tif}), while Ref.~\cite{Ma:2023vvr} 
finds
a very large Bayes factor of $\sim 600$ in favor of the overtone presence.

These conflicting results highlight the difficulty of analyzing the ringdown of existing events, due to their low signal-to-noise ratio (SNR) and to subtleties in the statistical methods employed. Therefore, current data require using robust statistical techniques that can deliver guaranteed coverage even in the presence of possible noise artifacts.
Moreover, although future detectors such as LISA, ET and CE will reach SNRs up to hundreds of times larger than GW150914, and will be guaranteed to observe more than one QNM~\cite{Berti:2016lat}, the statistical properties of the noise will be more complicated due to sources overlap, which will likely make a classic Bayesian analysis difficult if not unworkable.

Simulation-based inference (SBI), which 
dispenses with the need for a likelihood, provides a powerful alternative to the traditional Bayesian techniques employed in all the aforementioned ringdown analyses to date, and can be easily extended to include non-Gaussianity and non-stationarity in the noise.
Importantly, the amortized (i.e.~pre-trained) nature of SBI can be exploited, as we show here, to construct coverage maps leading to calibrated posteriors with guaranteed (i.e.~exact) coverage --- unfeasible with other methods. 
This aspect is especially important for the ringdown analysis of LVK events, due to their low ringdown SNRs.

SBI in the context of GW physics was introduced by Ref.~\cite{Chua:2019wwt} and then improved in Refs.~\cite{Green:2020hst, Green:2020dnx, Dax:2021tsq, Dax:2021myb, Dax:2022pxd, Wildberger:2022agw, Bhardwaj:2023xph}, resulting in  state-of-the-art parameter estimation for LVK events. However,  these works analyzed the entire inspiral-merger-ringdown GW signal, which has significantly larger SNR than the ringdown alone.

In this paper, we tackle the subtle problem of inferring the parameters of the remmant BH using the ringdown data alone. We do so by 
using an SBI pipeline relying on masked autoregressive flows, which we train against simulations and which we 
apply to get neural posterior estimates for the controversial case of the GW150914 ringdown.
Unlike previous SBI implementations in GW physics, 
 we also calibrate our posteriors by checking their coverage
against injected simulations
(i.e. we check  the frequentist probability that the posteriors include the injected parameter values). The resulting calibrated pipeline therefore produces posteriors with {\it guaranteed} confidence regions. When applied to GW150914, our calibrated posteriors show only mild evidence of the first overtone of the $\ell=m=2$ mode, with zero amplitude for the latter falling within the 95\% confidence region of the (marginalized) posteriors.

 \section{Method}
 
 \subsection{The model}
 \label{WF}
 
The detector response to a GW  is given by $h(t) = F_+ h_+ + F_\times h_\times$, where $F_{+,\times}(\alpha, \delta, \psi)$ are the antenna pattern functions (with $\alpha, \delta, \psi$ the right ascension, declination and polarization angles) and $h_{+,\times}$ the two GW polarizations\footnote{In general, one would also have to include the detector transfer function, but that is $\approx 1$ at the frequencies of GW150914 \cite{Rakhmanov:2008is}.}. The latter can be decomposed as $h_+ - i h_\times = \sum_{\ell m} h_{\ell m}(t) {}_{-2}Y_{\ell m }(\iota, \phi)$, where ${}_{-2}Y_{\ell m }(\iota, \phi)$ are the spin weighted spherical harmonics, which depend on the direction angles $(\iota, \phi)$ of the source relative to the detector. 
In the ringdown, each multipole component can be expressed as a linear combination of QNMs as
\be
h_{\ell m}(t) = \sum_{n=0}^\infty A_{\ell m n } e^{i \, \phi_{\ell m n} - \left(i \, \omega_{\ell m n} + \tau_{\ell m n}^{-1}\right)(t-t_0)} \,,
\ee
where we have neglected the mixing between modes 
with the same $m$ but different $\ell$, and we have assumed that all modes start at the same time $t_0$, which we fix to the peak of the amplitude of the inspiral-merger-ringdown signal. We also assume  progenitor spins aligned with the binary orbital angular momentum, and therefore the waveforms satisfy $h_{\ell m} = h_{\ell -m}^*$.
The integer $n$ labels the fundamental mode ($n=0$) and its overtones ($n=1,2,\dots$), in order of decreasing damping times.
The QNM frequencies $\omega_{\ell m n}$ and damping times $\tau_{\ell m n}$ are functions of the remnant mass $M_f$ and dimensionless spin $\chi_f$ only, and can be computed using perturbation theory \cite{Berti:2009kk}; we employ here the tabulated values given in Ref.~\cite{Berti}.
The amplitudes $A_{\ell m n }$ and phases $\phi_{\ell m n}$, describing how each mode is excited during the merger, cannot be computed in perturbation theory, and instead need to be estimated from the data. Following \cite{Isi:2019aib, Cotesta:2022pci}, in our analysis we consider only the fundamental mode ($n=0$) and the first overtone ($n=1$) of the dominant $\ell=m=2$  harmonic, since other modes are expected to be subdominant \cite{Buonanno:2006ui, Berti:2007fi}. Our parameter set is thus given by $\bm \theta = \{ M_f, \chi_f, A_{2 2 0 }, \phi_{2 2 0}, A_{2 2 1 }, \phi_{2 2 1}\}$.

We assume uncorrelated Gaussian noise in the frequency domain, described by the appropriate power spectral density (PSD) and, after transforming to the time domain, we inject the ringdown waveform described above. Full details are provided in Appendix~\ref{details}.
 
\subsection{Simulation-based inference}
 
SBI (cf. \cite{Cranmer:2019eaq} for a recent review) is a modern approach to Approximate Bayesian Computation that uses simulations from a model to estimate an approximate likelihood~\cite{2018arXiv180507226P}, likelihood-to-evidence ratios~\cite{Hermans:2019ioj, Delaunoy:2022tbl, Miller:2022haf, Karchev:2022xyn}, or posteriors~\cite{2016arXiv160506376P, 2019arXiv190507488G}. 

\subsubsection{Sequential Neural Posterior Estimation with Masked Autoregressive Flows}

 Given a series of pairs $\{\bm{\theta}_n, \bm{x}_n\}$, $n=1,\dots, N$, where $\{ \bm{\theta}_n \}$ is a set of parameters drawn from the prior $\pi(\bm{\theta})$ and $\{ \bm{x}_n \}$ is the corresponding set of simulated data, we wish to learn the posterior $p(\bm{\theta} | \bm{x}_*)$, where $\bm{x}_*$ are the real data. To this end, we train a neural network $F$, with weights $\psi$, to learn the mapping between data $\bm{x}$ and the parameters $\phi$ describing a family of conditional densities $q_\phi(\bm{\theta} | \bm{x})$ that act as parametric approximation to the posterior. Ref. \cite{2016arXiv160506376P} shows that maximizing the probability of $\prod_n q_\phi(\bm{\theta}_n | \bm{x}_n)$ with respect to $\phi$, for $N \to \infty$, leads to $q_\phi(\bm{\theta} | \bm{x}) \propto p(\bm{\theta} | \bm{x})$. This is achieved by minimizing the cross-entropy loss ${\cal L(\psi)}=-\sum_{n=1}^N \text{log}\,q_{\phi(\bm{x}_n,\psi)}(\bm{\theta}_n)$.

For the density estimation, we rely on a type of normalizing flow that stacks a series of autoregressive models, referred to as Masked Autoregressive Flow (MAF) \cite{2017arXiv170507057P}.
Autoregressive models \cite{2016arXiv160502226U} decompose the joint posterior density as a product of one-dimensional conditionals as $p(\bm{\theta}|\bm{x}) = \prod_i q_{\phi_i}(\theta_i | \bm{\theta}_{1:i-1}, \bm{x})$. Each conditional is modeled as a parametric density whose parameters are a function of a hidden state $\phi_i$, which in turn is a function of the previous hidden state $\phi_{i-1}$ and the current input variable $\theta_i$. MAF uses an autoregressive model called Masked Autoencoder for Distribution Estimation (MADE) \cite{2015arXiv150203509G}, which enforces the autoregressive property (i.e. the ordering of the variables) with a single forward pass,  dropping out connections by multiplying the weight matrices of a fully-connected autoencoder with suitably constructed binary masks.
An autoregressive model can be equivalently interpreted as a normalizing flow (see \cite{2017arXiv170507057P} for details).

Normalizing flows \cite{2015arXiv150505770J} transform a base density $q(\bm{\vartheta})$ (usually a standard Gaussian) into the target density $p(\bm{\theta}| {\bm{x}})$ via an invertible transformation $\bm{\theta} = f(\bm{\vartheta})$ with tractable Jacobian:
\be
p(\bm{\theta} | \bm{x}) = q\left(f^{-1}(\bm{\theta})\right) \left\vert \text{det}\left( \frac{\partial f^{-1}}{\partial \bm{\theta}} \right) \right\vert \,.
\ee
The flow is usually obtained from composing multiple such invertible transformations.
In practice, MAF uses MADE with Gaussian conditionals as the building layer of the flow.

\subsubsection{Priors truncation scheme}
\label{Truncated_priors}

The performance of the neural network crucially depends on the amount of relevant training examples it receives.
For this reason, several iterative refinements of the posterior estimate have been proposed, and go under the name of sequential neural density estimation.
These approaches focus the training on regions of parameter space compatible with a specific target dataset by successively restricting the prior range from which training examples are generated~\cite{2016arXiv160506376P, 2017arXiv171101861L, 2019arXiv190507488G}. We employ a truncation scheme \cite{Karchev:2022xyn, Miller:2021hys, Miller:2022shs, Miller:2020hua} that does not modify the shape of the prior distribution but merely restricts its support, thus avoiding the need for later adjustments to the posterior density estimate. At each stage, we truncate the uniform prior (independently for each parameter)
by excluding the regions where the posterior density has negligible support.
 Specifically, we restrict the prior by keeping only the highest probability density region that contains a $4\sigma$ volume of the approximate posterior (evaluated for the target data).
After each truncation, a new network is initialized and trained on samples generated with the newly constrained priors.
We stop truncating when none of the parameter ranges shrinks by more than a factor of two. 

\subsubsection{Validation and calibration}
\label{calibration}

The truncation scheme described above provides a locally amortized inference, i.e. within the prior region after the last stage of truncation. This enables calibration of the resulting posterior regions to produce regions with guaranteed exact frequentist coverage \cite{Karchev:2022xyn}---a desirable property to ensure robustness of the scientific conclusions drawn from the inference. 

The {\it credibility} $\gamma$ of a parameter value $\bm{\theta}_0$ is the approximate posterior probability enclosed by the region where the approximate posterior density, $q(\bm{\theta}|\bm{x})$, is higher than at $\bm{\theta}_0$, i.e.
\be
\gamma(\bm{\theta}_0, \bm{x}) \equiv \mkern-36mu\mkern-36mu \intop_{\qquad\qquad q(\bm{\theta}|\bm{x})>q(\bm{\theta}_0|\bm{x})} \mkern-36mu\mkern-36mu q(\bm{\theta}|\bm{x}) \, \text{d} \bm{\theta} \,.
\ee
By repeatedly simulating data $\bm{x}_0$ from $\bm{\theta}_0$ and deriving $\gamma(\bm{\theta}_0, \bm{x}_0)$, the cumulative distribution of these credibilities gives the {\it empirical coverage} $\mathcal{C}$ of the approximate posteriors, i.e. $\mathcal{C}(\gamma) \equiv \int_0^\gamma {\cal P}(\gamma') \text{d}\gamma'$.
The latter represents the frequency with which regions of different credibility include (cover) $\bm{\theta}_0$.
If $\mathcal{C}>\gamma$, the approximation is said to be conservative (it covers more frequently than its credibility), while if $\mathcal{C}<\gamma$ it is said to be under-covering.

Bayesian validation consists of checking whether the empirical coverage matches the credibility calculated with the exact posterior,
by examining deviations from the diagonal line in a Bayesian P–P plot, on average across the prior range.
However, a diagonal Bayesian P–P plot is not a sufficient condition because of the averaging:
conservative regions can compensate under-covered ones.
Alternatively, if data are simulated at fixed parameter values, one can build a frequentist validation plot for each point across the support of the prior. From these plots, one can derive a map of {\it required credibility} $\hat\gamma$ for any given confidence level $\tilde\gamma$, as the credibility that has frequentist coverage $\tilde\gamma$ (i.e. $\mathcal{C}(\hat\gamma)=\tilde\gamma$). In other words, $\hat\gamma$ is the $\tilde\gamma^\text{th}$ quantile of $\mathcal{C}$.
From this map, one can construct frequentist confidence regions with {\it guaranteed exact}  coverage: for a given observation $\bm{x}$,  one has to include in the region the parameters $\bm{\theta}_0$ 
for which $\gamma(\bm{\theta}_0, \bm{x}) < \hat\gamma(\bm{\theta}_0, \tilde\gamma)$.
This procedure will be used to produce the calibrated confidence regions of Fig.~\ref{fig} (shaded areas).

\section{Results and discussion}

\begin{figure*}[htp]
  \centering
  \subfigure{\includegraphics[scale=0.5]{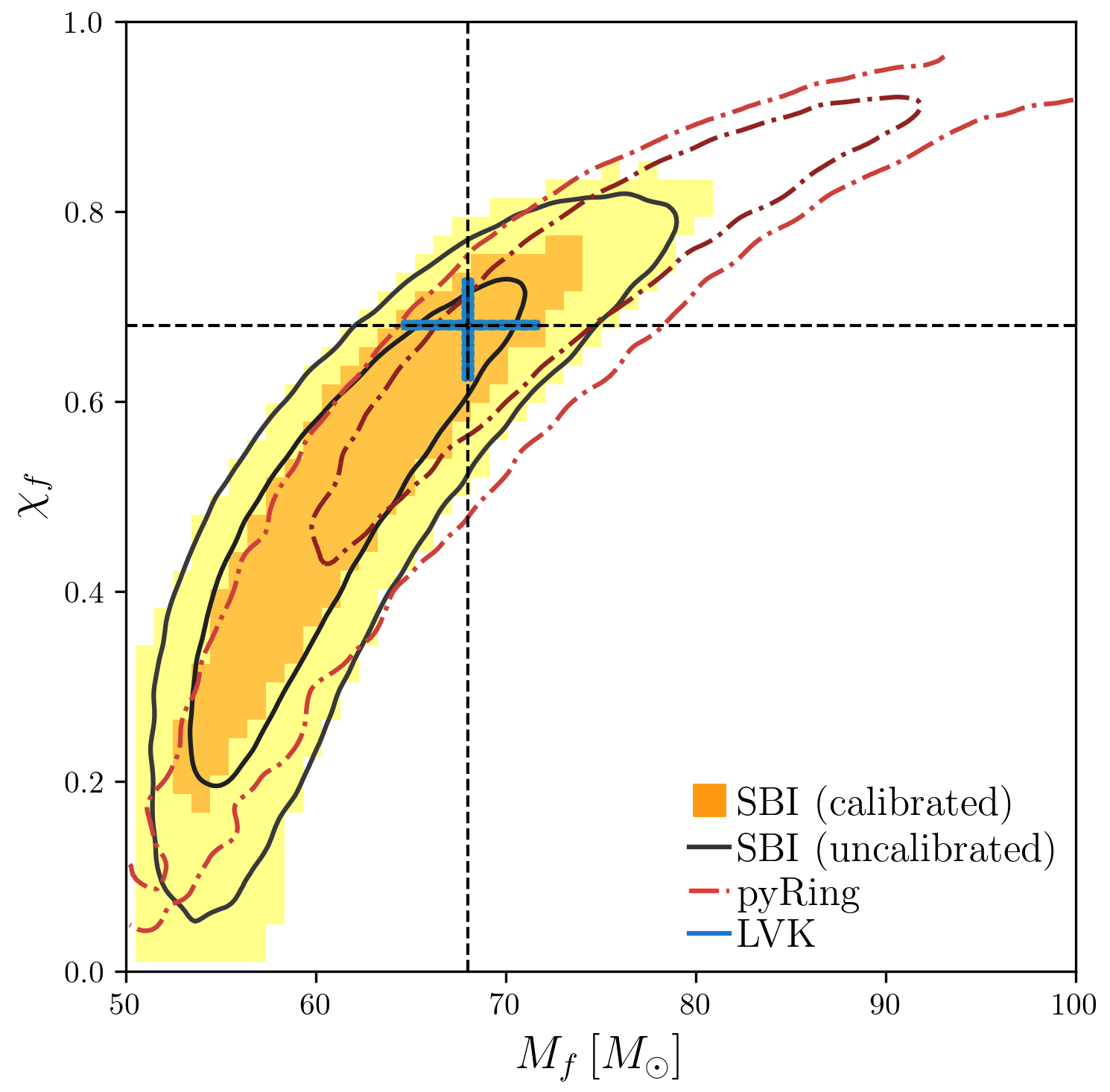}}\quad
  \subfigure{\includegraphics[scale=0.56]{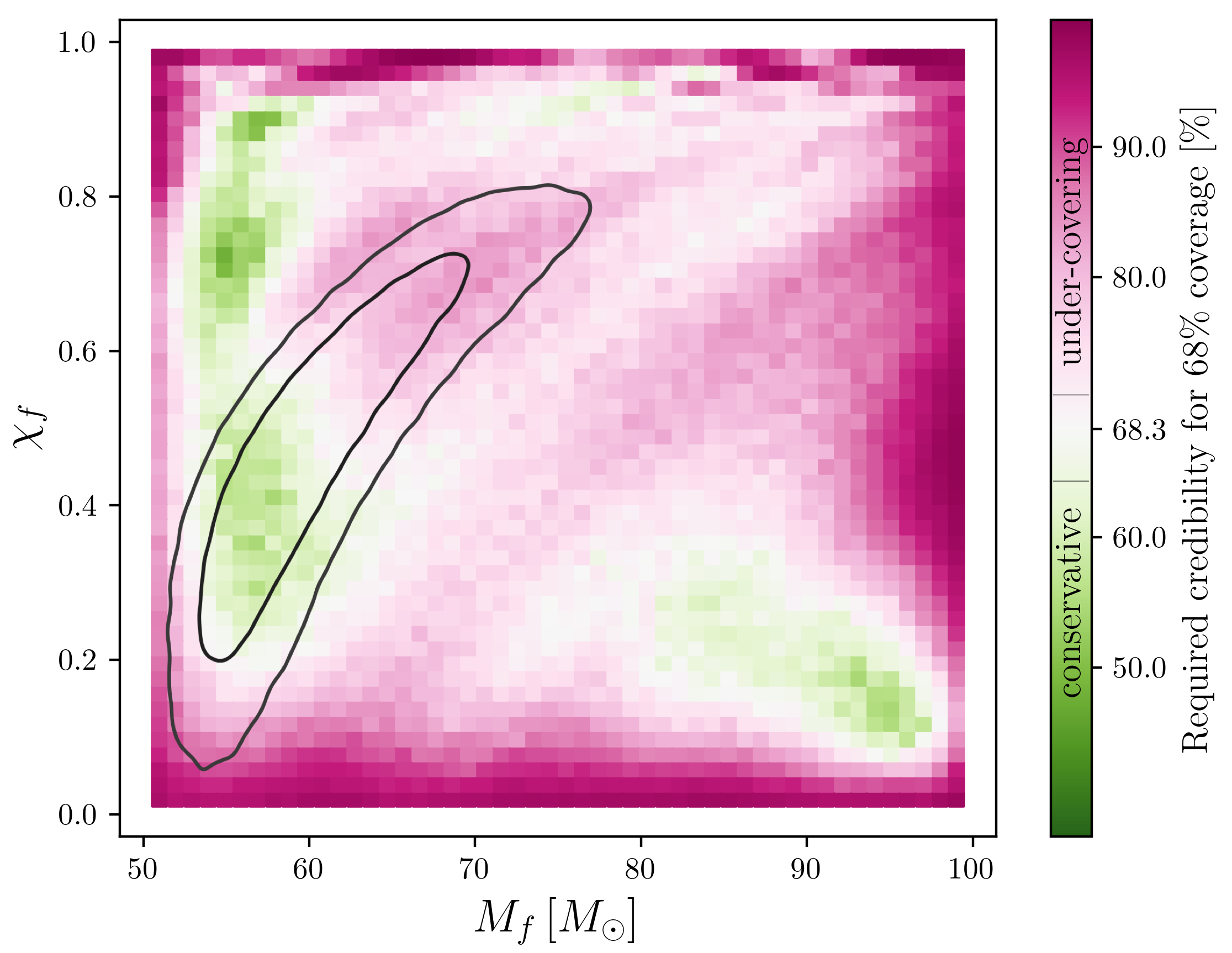}} \\
  \subfigure{\includegraphics[scale=0.5]{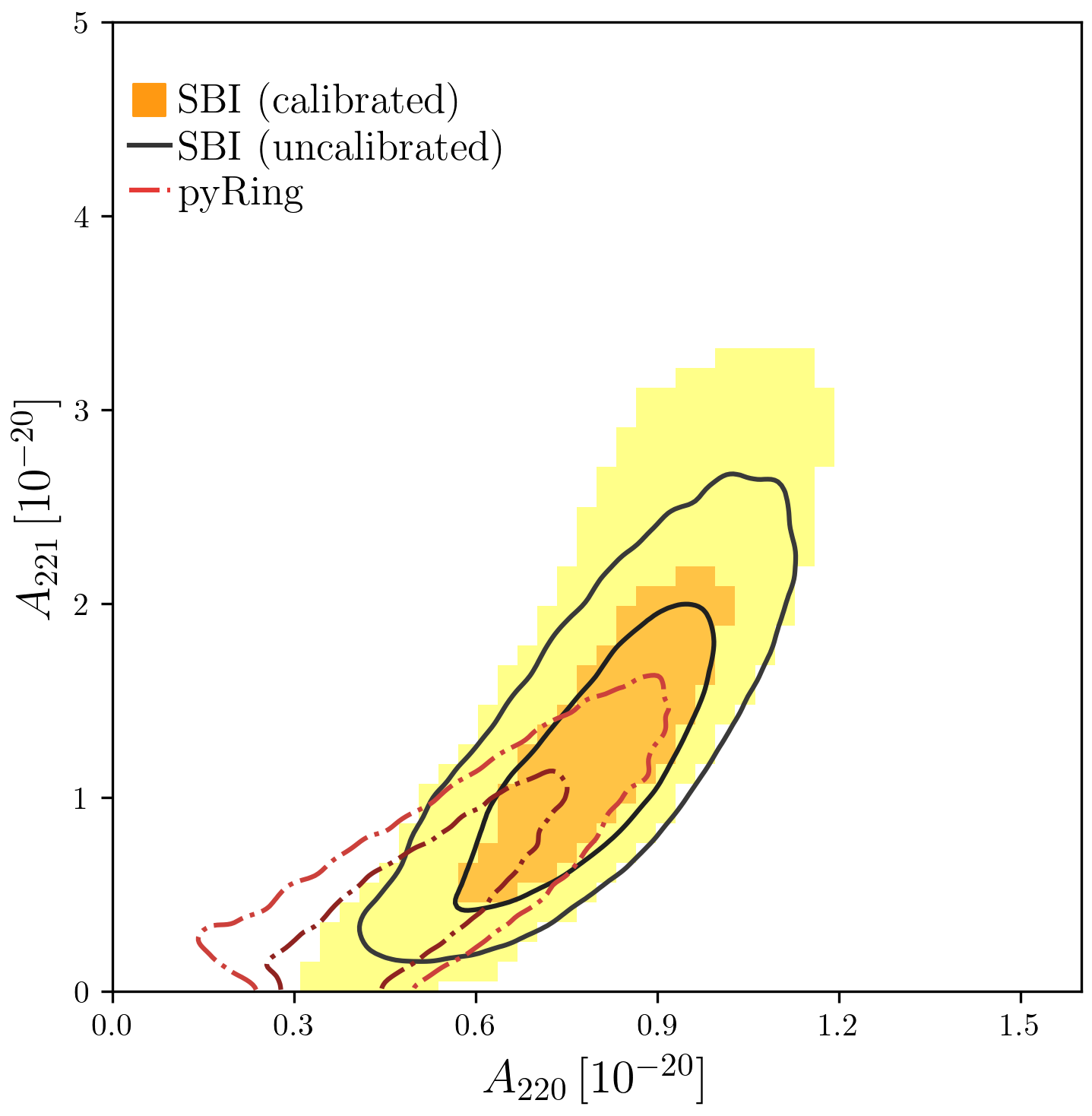}}\quad\,\,\,
  \subfigure{\includegraphics[scale=0.56]{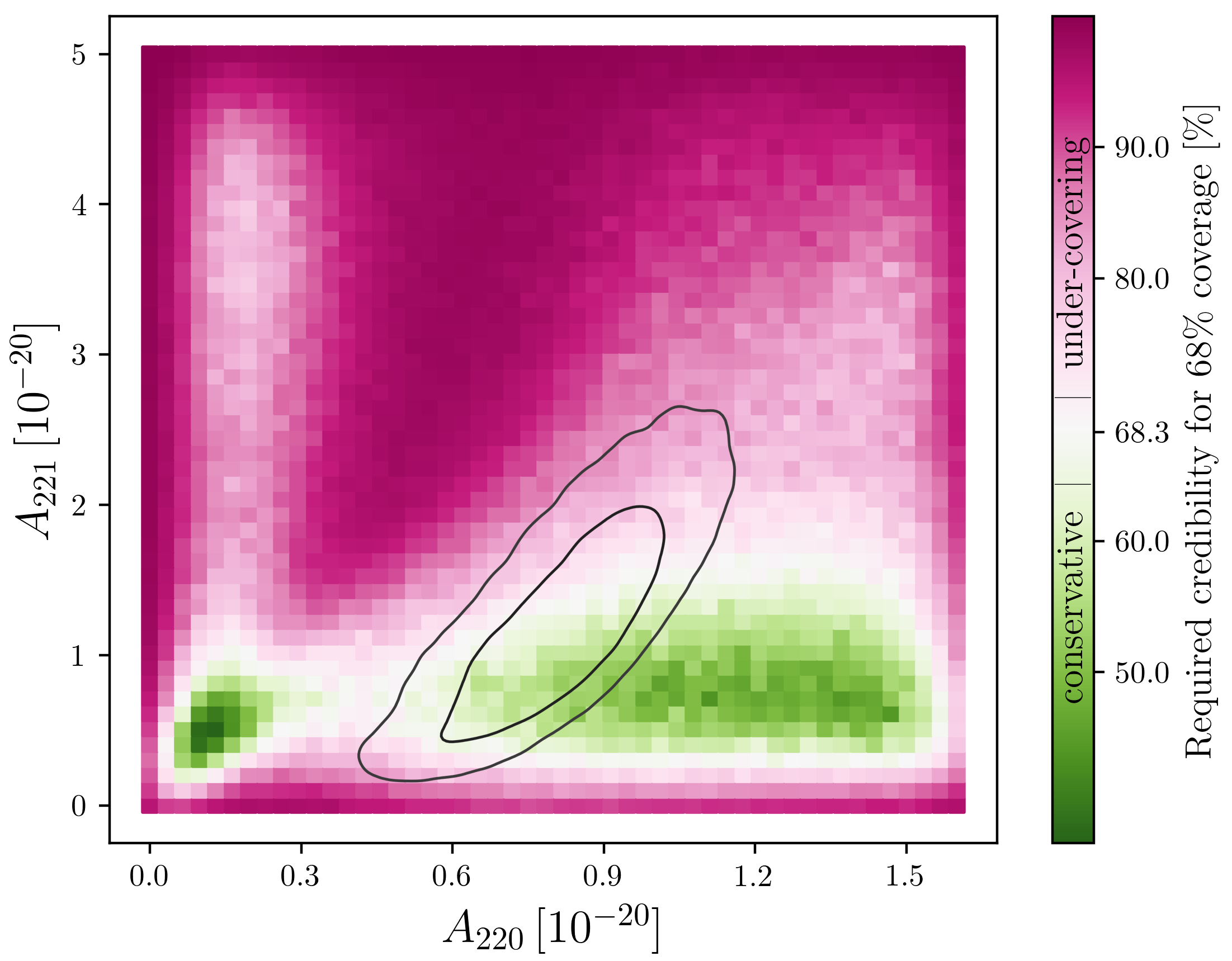}}
  \caption{Results for the 
GW150914 ringdown analysis. Left panels: marginal posterior 68\%, 95\% regions from the uncalibrated simulation-based analysis (solid/black) and calibrated regions with guaranteed exact coverage (68\% orange, 95\% yellow) for mass and spin (top panel) and QNM amplitudes (bottom panel). The red/dot-dashed contours are from a likelihood-based analysis. Right panels: maps of required credibility (for 68\% guaranteed coverage) used to obtain the calibrated regions on the left. The black contours are as in the left panel, for reference. }
  \label{fig}
\end{figure*}

In the left panels of Fig.~\ref{fig}, we show results for the 
GW150914 ringdown (starting from the peak of the amplitude). Shaded areas represent the $68\%$ and $95\%$ calibrated (i.e. guaranteed-coverage) credible regions for the mass and spin (upper plot) and for the amplitude of the fundamental mode and first overtone (lower plot).
All the other parameters are marginalized on.
The solid lines
delimit the $68\%$ and $95\%$ credible regions from the approximate posterior 
obtained from the networks before calibration, while the dot dashed lines show the same credible regions inferred with a likelihood-based 
Bayesian analysis, i.e. with {\tt pyRing} \cite{pyring, Carullo:2019flw, LIGOScientific:2020tif}, 
the official  ringdown analysis tool of the LVK collaboration.
Uncalibrated marginal distributions for the full six-dimensional posteriors are shown in Appendix~\ref{full}, Fig.~\ref{figapp}.
The blue thick cross in the upper plot shows
 the median values (with $90\%$ credible intervals represented by the arm-lengths) 
 of the remnant mass and spin as estimated by the LVK collaboration
 from the full inspiral-merger-ringdown signal,
 using two precessing waveform models (EOBNR and IMRPhenom) and averaging the corresponding posteriors with equal weights~\cite{LIGOScientific:2016wkq}. All results are for a joint analysis of the LIGO Hanford and LIGO Livingstone data (see Appendix~\ref{details} for details).

 In the right panels, we show the maps of the required credibility for 68\% exact coverage resulting from the amortized approximate posteriors. At each discrete parameter value (at the center of each pixel shown), we fix all other parameters to the most probable  values obtained from the approximate posteriors of the uncalibrated network for the GW150914 ringdown (given in Appendix~\ref{full}, Fig.~\ref{figapp}).
We simulate the signal and $10^3$  realizations of the LIGO noise at each pixel. For each
 time series (signal plus noise) simulated in this way, 
we then derive the amortized approximate posterior and use it to compute 
the frequentist probability required to cover the injection's true mass and spin (top) or amplitudes (bottom) at 68\% and 95\% confidence level. The required credibility for 68\% coverage is shown by the color code. For instance, a color code corresponding to say 80\% (50\%) means that for an injection with parameters in that pixel, the nominal 80\% (50\%) confidence regions computed from the approximate posteriors actually contain the true parameter values
68\% of the times, i.e. the nominal confidence regions produced by the uncalibrated network are too small (large), signaling under-covering (conservative) posteriors.

Even though our approximate posterior exhibits over- and under-coverage in some regions (for $68\%$ confidence), this is actually the result of rather small inaccuracies in the size of the inferred posterior. Indeed, as can be seen, the calibrated $68\%$ contours only slightly depart from the approximate ones.
For mass and spin, this is also true for the $95\%$ contour.
As for the amplitudes, the approximate posterior distribution tends to be under-covering (for $95\%$ coverage) for higher values of $A_{221}$, which results in a more sizeable enlargement of the $2\sigma$ contour in that direction.

We stress again that this coverage calibration procedure is enabled by the locally amortized nature of our SBI technique --- being computationally unfeasible with traditional Bayesian methods. Fig.~\ref{fig} requires $5 \times 10^6$ full simulated inferences (each taking less than a second with our trained network), which would require $\sim 5 \times 10^{11}$ likelihood evaluation with {\tt pyRing}. 
Most importantly, the calibration procedure leads to regions with guaranteed coverage. The procedure, in general, tends to enlarge the contours in the under-covered regions (left panels of Fig.~\ref{fig}). This can intuitively be understood because the network is overly confident in the under-covered region (therefore resulting in artificially small contours).

Interestingly, while our approximate posteriors seem to provide some support for the presence of the first overtone in the data (excluding $A_{221}=0$ at $95$\% level), after  calibration $A_{221}=0$ lies within the $95$\% credibility region. Since in this analysis we considered the GW150914 data starting at the amplitude peak (i.e.~we put ourselves in the most favorable scenario for detecting the first overtone \cite{Baibhav:2023clw}), we conclude that our guaranteed-coverage results yield
 only mild evidence ($\sim 2\sigma$) for the $\ell=m=2$, $n=1$ mode in the data\footnote{We have also performed analyses at other 12 different starting times, namely every $0.23 \text{ms}$ in the range $t_{\text{peak}} \pm 1.38 \text{ms}$, and found the same behavior as in Ref.~\cite{Cotesta:2022pci}: the approximate posteriors provide increasing support for the presence of the first overtone for earlier starting times, whereas decreasing support for later starting times.}. 
Our findings therefore cast doubts on the usefulness of GW150914 to test the no-hair theorem of General Relativity, although such tests will definitely become feasible with next generation (ground- and space-based) detectors~\cite{Berti:2016lat}.

Our results can be extended in several directions. As mentioned, SBI techniques only require forward simulations, and can therefore account for non-Gaussian/non-stationary noise features provided that those can be simulated (or sampled from chunks of data where no signal is present), thus improving the realism of the detector description. A powerful alternative to the flow-based neural posterior estimation used here is neural ratio estimation, which could be used to distinguish between a model with and without a first overtone. Finally, population-level inference for black hole binaries is another target ripe for exploration with SBI techniques, given the requirement to model the complex selection function of the detectors. We believe that SBI will play a prominent role in scientific exploitation of future GW data. 
  
\acknowledgments
\no We thank Gregorio Carullo, Roberto Cotesta, Vasco Gennari and Walter Del Pozzo for helpful discussions and correspondence about the {\tt pyRing} package, and Costantino Pacilio, Konstantin Karchev, Christoph Weniger for useful discussions.
MC and EB acknowledge support from the European Union’s H2020 ERC Consolidator Grant “GRavity from Astrophysical to Microscopic Scales” (Grant No. GRAMS-815673) and the EU Horizon 2020 Research and Innovation Programme under the Marie Sklodowska-Curie Grant Agreement No. 101007855. RT acknowledge co-funding from Next Generation EU, in the context of the National Recovery and Resilience Plan, Investment PE1 – Project FAIR ``Future Artificial Intelligence Research''. This resource was co-financed by the Next Generation EU [DM 1555 del 11.10.22]. RT is partially supported by the Fondazione ICSC, Spoke 3 ``Astrophysics and Cosmos Observations'', Piano Nazionale di Ripresa e Resilienza Project ID CN00000013 "Italian Research Center on High-Performance Computing, Big Data and Quantum Computing" funded by MUR Missione 4 Componente 2 Investimento 1.4: Potenziamento strutture di ricerca e creazione di ``campioni nazionali di R\&S (M4C2-19 )'' - Next Generation EU (NGEU).

\appendix

\section{Technical Details}
\label{details}

We use a sample rate of 4096 Hz and analyse 0.1 seconds of LIGO data for the GW150914 ringdown event.
We take as starting time at the LIGO Hanford detector $t_0=t_{\text{peak}}=1126259462.42323$ GPS, as estimated in \cite{Cotesta:2022pci}, and compute the time of arrival in LIGO Livingston with the {\tt LALSuite} library \cite{LAL}.
The antenna pattern functions are also provided by the {\tt LALSuite} library \cite{LAL}, and we fix the sky position of the source  as assumed in \cite{Isi:2019aib}, namely $\alpha = 1.95$ rad, $\delta = -1.27$ rad, $\psi = 0.82$ rad and inclination $\iota = \pi$ rad, $\phi=0$. 
We assume uniform priors on all parameters with $M_f \in [50, 100] \, M_\odot$, $\chi_f \in [0, 1]$, $A_{22n} \in [0, 5] \cdot 10^{-20}$ and $\phi_{22n} \in [0, 2\pi]$.

We perform our analysis in the frequency domain and apply a band-pass filter between 100 and 650 Hz. This reduced range of frequencies covers completely the ringdown signal for masses and spins in our prior space, and allows us to avoid any data compression.
To produce the training set coherently with the data that we analyze, we first generate the strain data in the time domain, and then we transform to the frequency domain. We generate random noise realizations from the PSDs provided by the LVK collaboration for 32 seconds of LIGO Hanford/Livingston data around GW150914 \cite{LVK}\footnote{An alternative, which results very useful when a reliable PSD is not available due to non-Gaussianity and/or non-stationarity in the data, is to use chunks of real noise taken from the data around the event.}, and we inject signals using  the model given in Sec.~\ref{WF}. After cutting the data 0.1s after the starting time, we apply a Tukey window function with $\alpha=1/70$ and perform a Real Fast Fourier Transform (RFFT) followed by whitening. Finally, we concatenate the real and imaginary parts of the RFFT to produce our 112 bin long dataset.
The very sharp window function that we use introduces contamination in the form of spectral leakage. However, since this happens coherently in the training set and in the observable, in our case this does not represent an issue. In other words, one can think of the RFFT as a linear transformation of the data consistently applied to both simulations and data.

We train one neural network for each of the two LIGO detectors, and  we perform the inference separately in each one. We then  combine the posteriors by rejection sampling.
To speed up the production of the required credibility maps as described in Sec.~\ref{calibration}, we also combine the two posteriors doing importance sampling through weighted Kernel Density Estimation (KDE). Namely, we perform KDE on the samples from the LIGO Hanford analysis and weigh them with the posterior probability obtained from the LIGO Livingston analysis.
We have checked this procedure to be equivalent to rejection sampling and sensibly faster.
We validate the posteriors at fixed parameter values with P-P plots on two-dimensional grids of 2500 points each, across the final truncated prior, using 1000 simulations per point. The resulting maps for 68\% coverage are given in the right panels of Fig.~\ref{fig}.

\section{Software and settings}

For simulation-based-inference we rely on the package {\tt sbi}~\cite{tejero-cantero2020sbi, sbi} (v0.21.0).
For training, we use the Adam optimizer \cite{Kingma:2014vow} with learning rate $5\cdot10^{-4}$.
In the truncation procedure, at each stage $5\cdot 10^6$ new simulations are generated on-the-fly (10\% of which are used for validation), and the training proceeds in batches of size 100.
In more details, every training set consists of $10^5$ signals (and corresponding parameters), each of them paired with 50 random realizations of the noise. 
We found this configuration to be optimal for our case: a larger number of  signals did not improve the inference, whereas providing multiple noise realizations for each element sensibly increased the learning capacity of the network.
We stop each training stage when the validation loss ceases to decrease for 25 consecutive epochs.
Each training epoch takes roughly 10 minutes on an NVIDIA A-100 GPU.

The training of the LIGO Hanford network proceeded over two stages of iterative refinements as described in Sec.~\ref{Truncated_priors}, the first lasting for 107 epochs and the second for 125 epochs. The LIGO Livingston network instead does not trigger any iterative refinement of the priors, signalling a lower SNR than in Hanford, as already known from standard Bayesian analyses. Here, the training proceeds over a single stage of 145 epochs.
We have checked that for synthetic data with higher SNR, the iterative refinement of the priors is very efficient, and scales roughly as one extra round every time the SNR doubles.
This prior truncation scheme will be extremely useful when applied to the analysis of the ringdown phase of LISA/ET/CE-like events.
Instead, the calibration procedure, described in Sec.~\ref{calibration}, seems unnecessary for higher SNRs: already at SNR five times larger than in GW150914 the required credibility maps match the exact coverage at very high precision level (few \%) all-over the (last) prior space and the calibrated regions coincide with the uncalibrated ones.

We use MAF with 5 autoregressive layers and a standard Gaussian as a base density, using the default order for the first layer (i.e. the layer that directly models the data) and reversing the order for each successive layer. Each of them contains two hidden layers of 120 units each. 

For the comparison with standard Bayesian inference, we employ the package {\tt pyRing} (v2.2.1) \cite{pyring}, which in turn relies on the nested sampling algorithm {\tt cpnest} (v0.11.5) \cite{cpnest}, and we used 4096 live points and 4096 maximum Markov Chain steps.
For the analysis we used the same framework as for sbi (with ACF estimated from the PSD), but the frequency window which in {\tt pyRing} was $20-2038 \text{Hz}$.

\section{Approximate posteriors for all parameters}
\label{full}

\begin{figure*}
\begin{center}
\includegraphics[width=0.7\textwidth]{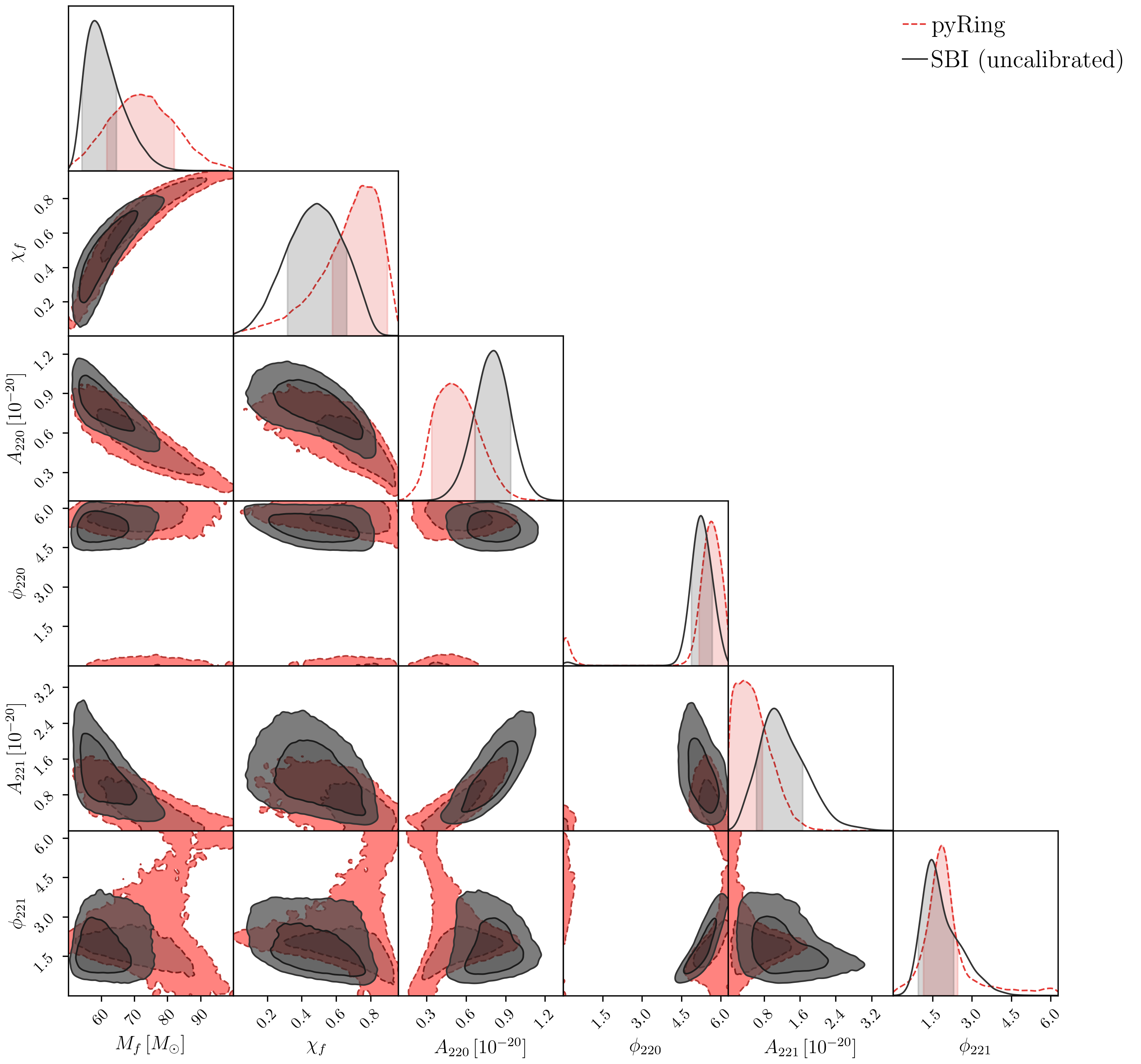}
\caption{68\% and 95\% marginal posteriors for the GW150914 ringdown analysis, from the uncalibrated simulation-based analysis (solid/black) and the likelihood-based analysis (red/dashed).}
\label{figapp}
\end{center}
\end{figure*}

In Fig.~\ref{figapp} we show the uncalibrated 68\% and 95\% marginal distributions for the full six-dimensional posteriors from the GW150914 ringdown analysis, together with the same credible regions inferred with {\tt pyRing} as a reference.

\bibliographystyle{utphys}
\bibliography{bibliography}

\end{document}